\documentclass[aps,preprint,groupedaddress,showpacs,preprintnumbers,amsmath,amssymb]{revtex4}
\usepackage{graphicx}
\newcommand{\rarr}{\rightarrow}

\newcommand{\jpsi}{J/\psi}
\newcommand{\jpsito}{J/\psi \rightarrow}
\newcommand{\psip}{\psi(2S)}
\newcommand{\psipto}{\psi(2S) \rightarrow }

\newcommand{\pp}{\pi^+\pi^-}

\newcommand{\ee}{e^+e^-}
\newcommand{\mm}{\mu^+\mu^-}

\newcommand{\kzs}{K^{0}_{S}}

\begin{document}

\title{\boldmath Measurement of the branching fractions of $\psipto 3(\pp)$
and $\jpsito 2(\pp)$}

\author{
M.~Ablikim$^{1}$,      J.~Z.~Bai$^{1}$,       Y.~Ban$^{11}$,
J.~G.~Bian$^{1}$,      X.~Cai$^{1}$,          J.~F.~Chang$^{1}$,
H.~F.~Chen$^{17}$,     H.~S.~Chen$^{1}$,      H.~X.~Chen$^{1}$,
J.~C.~Chen$^{1}$
% ...
\\(BES Collaboration)\\
$^{1}$ Institute of High Energy Physics, Beijing 100049,
People's Republic of China\\
% ...
$^{11}$ Peking University, Beijing 100871, People's Republic of China\\
% ...
$^{17}$ University of Science and Technology of China, Hefei
230026, People's Republic of China\\
% ...
}

\date{\today}

\begin{abstract}
  
  Using data samples collected at $\sqrt{s} = 3.686$~GeV and
  $3.650$~GeV by the BESII detector at the BEPC, the branching
  fraction of $\psipto 3(\pp)$ is measured to be $[4.83\pm 0.38
  ~(\hbox{stat}) \pm 0.69 ~(\hbox{syst})] \times 10^{-4}$, and the
  relative branching fraction of $\jpsito 2(\pp)$ to that of $\jpsito
  \mm$ is measured to be $[5.86\pm 0.19 ~(\hbox{stat}) \pm 0.39
  ~(\hbox{syst})] \%$ via $\psipto \pp\jpsi, \jpsito 2(\pp)$. The
  electromagnetic form factor of $3(\pp)$ is determined to be $0.21
  \pm 0.02$ and $0.20 \pm 0.01$ at $\sqrt{s} = 3.686$~GeV and
  $3.650$~GeV, respectively.

\end{abstract}


\maketitle

\section{Introduction}

Strong decays of $\psip$ to $3(\pp)$ are suppressed, since the
reaction violates G-parity conservation. In $\ee$ colliding beam
experiments, $3(\pp)$ may also be produced by $\ee \rarr \gamma^{*}
\rarr 3(\pp)$ (called the ``continuum process'' hereafter). It is
expected that the continuum contribution is large and may contribute
around 60\% of the $3(\pp)$ events at the $\psip$ energy. This
contribution must be removed in determining $\mathcal{B}(\psipto
3(\pp))$, as has been described for the $\psipto \pp$ decay
mode~\cite{ff}.

In this analysis, data samples at the $\psip$ peak ($\sqrt{s}=3.686$
GeV) and off-resonance ($\sqrt{s}=3.650$~GeV) are used.  The continuum
contribution at the $\psip$ peak is estimated using the off-resonance
sample and subtracted to obtain a model independent measurement of the
$\psipto 3(\pp)$ branching fraction. We also use the samples to obtain
the $3(\pp)$ electromagnetic form factor which allows us to calculate
the branching fraction based on the theoretical assumption described
in Ref.~\cite{ff}.

There is a big contribution from $\psipto \pp \jpsi, \jpsito 2(\pp)$
in our $\psipto 3(\pp)$ sample. This process allows us to measure
the branching fraction of $\jpsito 2(\pp)$. The advantage of this
method is that we need not subtract the continuum contribution for
this process.

The existing branching fraction measurement of $\psipto 3(\pp)$ was
done by the Mark-I experiment \cite{mark1psp} based on $(9\pm 5)$
candidate events. The branching fraction of $\jpsito 2(\pp)$ was also
measured by Mark-I~\cite{mark1jpsi} with $(76 \pm 9)$ events observed,
and there is also a recent result for this decay reported by the BABAR
experiment \cite{babar}.

\section{The BES Experiment}

The data used for this analysis are taken with the updated Beijing
Spectrometer (BESII) detector at the Beijing Electron-Positron
Collider (BEPC) storage ring. The $\psip$ data are taken at $\sqrt{s}
= 3.686$~GeV with a luminosity of $\mathcal{L}_{3.686} = (19.72 \pm
0.86)~\hbox{pb}^{-1}$~\cite{lum} measured with large angle Bhabha
events. The number of $\psip$ events is $N^{tot}_{\psip} = (14.0 \pm
0.6)\times 10^6$~\cite{moxh} as determined from inclusive hadrons.
The continuum data are taken at $\sqrt{s} = 3.650$~GeV, and the
corresponding luminosity is $\mathcal{L}_{3.650}=(6.42 \pm
0.24)~\hbox{pb}^{-1}$~\cite{lum}. The ratio of the two luminosities is
$\mathcal{L}_{3.686}/\mathcal{L}_{3.650} = 3.07 \pm 0.09$.

The BESII detector is a conventional solenoidal magnet detector that
is described in detail in Refs.~\cite{bes,bes2}. A 12-layer vertex
chamber (VC) surrounding the beam pipe provides trigger and track
information. A forty-layer main drift chamber (MDC), located radially
outside the VC, provides trajectory and energy loss ($dE/dx$)
information for charged tracks over $85\%$ of the total solid angle.
The momentum resolution is $\sigma _p/p = 0.017 \sqrt{1+p^2}$ ($p$ in
$\hbox{\rm GeV}/c$), and the $dE/dx$ resolution for hadron tracks is
$\sim 8\%$.  An array of 48 scintillation counters surrounding the MDC
measures the time-of-flight (TOF) of charged tracks with a resolution
of $\sim 200$ ps for hadrons.  Radially outside the TOF system is a 12
radiation length, lead-gas barrel shower counter (BSC).  This measures
the energies of electrons and photons over $\sim 80\%$ of the total
solid angle with an energy resolution of $\sigma_E/E=22\%/\sqrt{E}$
($E$ in GeV).  Outside of the solenoidal coil, which provides a
0.4~Tesla magnetic field over the tracking volume, is an iron flux
return that is instrumented with three double layers of counters that
identify muons of momentum greater than 0.5~GeV/$c$.

A GEANT3 based Monte Carlo (MC) program with detailed consideration of
detector performance (such as dead electronic channels) is used to
simulate the BESII detector. The consistency between data and Monte
Carlo has been carefully checked in many high purity physics channels,
and the agreement is quite reasonable~\cite{simbes}.

In generating MC samples, initial state radiation is included, and
$1/s$ or $1/s^2$ dependent form factors are assumed where required. MC
samples of $\psipto \pp \jpsi$, $\jpsito X$ are generated with the
correct $\pp$ mass distribution \cite{jpsitoll}, and $\psipto
\pp \jpsi$, $\jpsito \mm$ is generated with the correct $\mm$ angle
distribution. Other samples are generated according to phase space.

\section{\boldmath Measurement of $\psipto 3(\pp)$}

\subsection{Event Selection}

Six charged tracks with net charge zero are required. Each charged
track, reconstructed using hits in the MDC, must have a good helix fit
in order to ensure a correct error matrix in the kinematic fit.
All six tracks are required to (1) originate from the beam
intersection region, i.e.  $\sqrt{V_x^2+V_y^2}<2$~cm and
$|V_z|<20$~cm, where $V_x$, $V_y$, and $V_z$ are the $x$, $y$, and $z$
coordinates of the point of closest approach to the beam axis, and (2)
have $|\cos\theta|\leq 0.8$, where $\theta$ is the polar angle of the
track.

A four constraint kinematic fit is performed with the six charged tracks
assuming all of them to be pions. If the confidence level of the
fit is greater than 1\%, the event is categorized as $\psipto 3(\pp)$.

\begin{figure}
  \includegraphics[height=8cm]{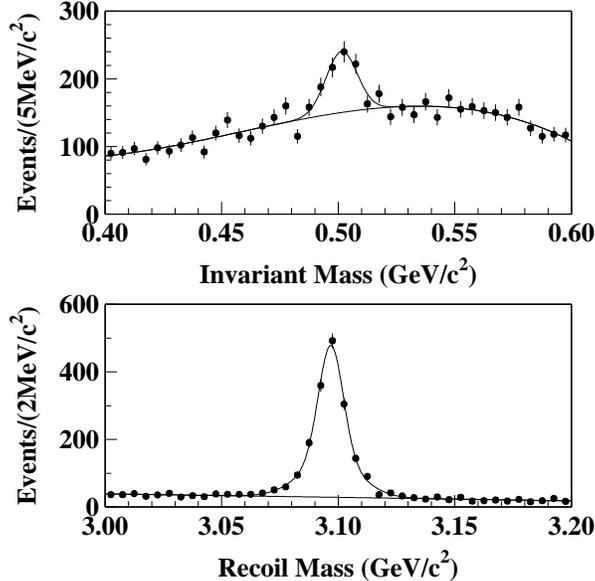}
\caption{\label{fig:mpp} The $\pp$ invariant and recoil mass
  distributions of $\psipto 3(\pp)$ candidates at $\sqrt{s}=3.686$
  GeV. The events with two $\pp$ pairs within
  $(0.47,0.53)$~GeV$/c^{2}$ are removed as $\kzs\kzs$ background. While
  the events with a $\pp$ pair within $(3.06,3.14)$~GeV$/c^{2}$ are
  removed as $\psipto\pp\jpsi$ background.}
\end{figure}

Fig. \ref{fig:mpp} shows the invariant and recoil mass distributions
of $\pp$ pairs. If the recoil mass of any $\pp$ pair is between 3.06
and 3.14 GeV$/c^{2}$, the event is considered a $\psipto \pp \jpsi$,
$\jpsito 2(\pp)$ candidate and removed. If the masses of any two $\pp$ pairs
is between 0.47 and 0.53 GeV$/c^{2}$, the event is considered  as
$\psipto$ $\kzs\kzs\pp$, $\kzs \rarr \pp$ background and removed.

Applying these criteria to the data collected at $\sqrt{s} =
3.686$~GeV, 670 events survive, while for the data collected at
$\sqrt{s} = 3.650$~GeV, 71 events remain. The efficiencies of these
criteria are $\varepsilon_{\psip} = 6.8\%$ for $\psipto 3(\pp)$ and
$\varepsilon_{cont} = 3.8\%$ for $\ee \rarr 3(\pp)$. The lower
$\varepsilon_{cont}$ results from the initial state radiation
correction (the maximum radiative photon energy is set to 0.7GeV) in
the generator, which reduces the center-of-mass energy for many events
generated. These events cannot survive the kinematic fit, which leads
to the lower $\varepsilon_{cont}$.

Remaining backgrounds in the $\psi(2S)$ sample include: (1)
residual  $\kzs \kzs$ or $\psipto\pp\jpsi$ events; (2) events with kaons
or electrons misidentified as pions, and (3) events with low energy
neutral tracks like $\pi^{0}$ or $\gamma$. Monte Carlo simulations
indicate that $N^{bg} = 4.9\pm 0.7$ events of these
backgrounds survive the above selection criteria, which will be
subtracted from the observed number of events in the calculation of
the branching fraction. Background remaining in the continuum data
sample is negligible.

\begin{figure}
  \includegraphics[height=7cm]{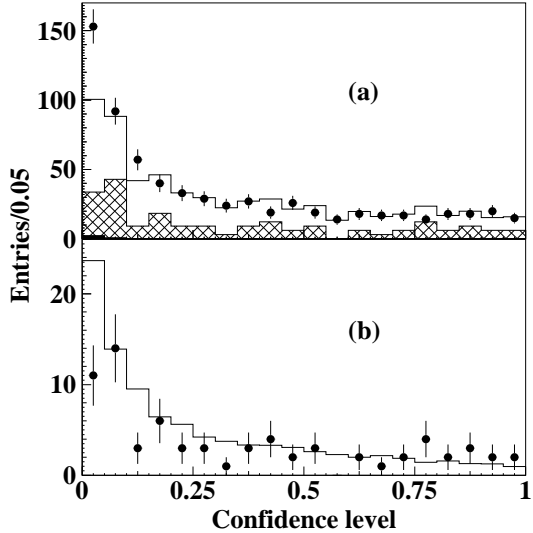}
\caption{\label{fig:x2} Confidence level distribution
  from the kinematic fitting. (a) is for $\psipto 3(\pp)$. The dots with
  error bars are data. The blank histogram is MC simulated
  signal plus the continuum contribution measured with $\sqrt{s} =
  3.650$ GeV data (hatched histogram) and the MC simulated background
  events (dark shaded histogram), after proper normalization. (b) is
  for $\ee \rarr 3(\pp)$ measured with $\sqrt{s} = 3.650$ GeV data.
  The dots with error bars are data, while the blank histogram is MC
  simulation. }
\end{figure}

Fig.~\ref{fig:x2} shows the confidence level distribution from the
kinematic fitting.  The consistency between data and MC is
satisfactory except for the first few bins in both
Figs.~\ref{fig:x2}(a) and (b).  Figure~\ref{fig:x2}(a) is similar to
the situation for $\jpsito 2(\pp)$ (see Fig.~\ref{fig:x2_jpsi}).
Taking into account the similarity between the figures and that we
expect little background in Fig.~\ref{fig:x2_jpsi}, we conclude that
the discrepancy is due to the simulation of the error matrix in the
track fitting rather than some unknown background, and this is
included in the systematic error analysis.

\subsection{Branching Fraction and Systematic Error}
To obtain the branching fraction of $\psipto 3(\pp)$, we have to
subtract from $N^{obs}_{3.686}$ the number of continuum events at
3.686GeV. This number is estimated as $N^{obs}_{3.650} \times f$,
where $f$ is the normalization factor:
\begin{eqnarray*}
  f=\frac{\displaystyle\mathcal{L}_{3.686}\times\sigma^{cont}_{3.686}}
  {\displaystyle\mathcal{L}_{3.650}\times\sigma^{cont}_{3.650}},
\end{eqnarray*}
where, $\sigma^{cont}$ is the Born order cross section of continuum
process, which is $s$ dependent and can be expressed in terms of the
$3(\pp)$ form factor $\mathcal{F}(s)$:
\begin{eqnarray}
  \sigma^{cont}(s)=\frac{4\pi\alpha^{2}}{3s}\times|\mathcal{F}(s)|^{2},
  \label{born1}
\end{eqnarray}
where $\alpha$ is the QED fine structure constant.
Assuming $\mathcal{F}(s) \propto 1/s$, we get:
\begin{eqnarray*}
  f&=&\frac{\displaystyle\mathcal{L}_{3.686}}
    {\displaystyle\mathcal{L}_{3.650}}
    \times
    (\frac{\displaystyle 3.650}{\displaystyle 3.686})^6\\
    &=&\frac{\displaystyle\mathcal{L}_{3.686}}
      {\displaystyle\mathcal{L}_{3.650}}
      \times 0.94.  
\end{eqnarray*}
The branching fraction of $\psipto 3(\pp)$ can be calculated as
\begin{eqnarray*}
  \mathcal{B}[\psipto 3(\pp)] &=&\frac{\displaystyle
    N^{obs}_{3.686}-N^{obs}_{3.650}\times f-N^{bg}}
  {\displaystyle \varepsilon_{\psip}\times N^{tot}_{\psip}}\\
%  &=&\frac{\displaystyle N^{obs}_{3.686}-N^{obs}_{3.650}\cdot
%    \frac{\mathcal{L}_{3.686}}{\mathcal{L}_{3.650}}\cdot 0.94-N^{bg}}
%  {\displaystyle \varepsilon_{\psip}\times N^{tot}_{\psip}}\\
  &=&(4.83 \pm 0.38 \pm 0.69) \times 10^{-4},
\end{eqnarray*}
where the first error is statistical and the second is systematic. The
values of the variables in the equation are listed in
Table~\ref{tab:val}.

\begin{table}
  \caption{\label{tab:val}Numbers used to calculate $\mathcal{B}[\psipto 3(\pp)]$.}
  \begin{tabular}{c|c|c|c|c|c}\hline
    $N^{obs}_{3.686}$&$N^{obs}_{3.650}$&$N^{bg}$&$N^{tot}_{\psip}$&
             $\varepsilon_{\psip}$&$\mathcal{L}_{3.686}/\mathcal{L}_{3.650}$\\\hline
    670&71&4.9&$1.4 \times 10^{7}$&6.8\%&3.07\\\hline
  \end{tabular}
\end{table}

The systematic errors come mainly from the MDC tracking, the
generator, the continuum subtraction, and the kinematic fit, as well
as the statistics of the MC samples.  The systematic error
contributions are listed in Table~\ref{tab:err1} and explained below.
The total systematic error is 15\%.

\begin{table}
  \caption{Summary of systematic errors.}
  \label{tab:err1}
  \begin{tabular}[t]{l|c}\hline
    Source       &systematic error(\%)\\\hline\hline
    MC statistics&0.7                  \\
    MDC tracking &12                   \\
    Kinematic Fit&4                    \\
    generator    &4                    \\
    continuum subtraction&4            \\
    total number of $\psip$&4          \\\hline
    Total        &15                   \\\hline
  \end{tabular}
\end{table}

\begin{enumerate}
  
\item The MDC tracking efficiency was measured using $\jpsi \rarr
  \Lambda \bar{\Lambda}$ and $\psipto \pp \jpsi,\jpsito \mm$ events. It is
  found that the MC simulation agrees with data within 1-2\% for each
  charged track. The largest difference is taken as a conservative
  estimation, and 12\% is quoted as the systematic error on the
  tracking efficiency for the channel of interest.

\item Fig.~\ref{fig:6pires} shows the $\pp$ mass distributions after
  applying all the selection criteria; a clear $\rho$ signal is
  observed in both the $\psip$ and the continuum samples. To estimate
  the efficiency difference between the data and MC simulation (pure
  phase space), we generate samples with different intermediate states
  in $\psipto 3(\pp)$, e.g., $\rho$, $\omega$, etc., and find the
  differences from pure phase space are at the 4\% and 5\% level for
  $\psip$ decays and continuum data, respectively, which are taken as
  the systematic errors from the generator.

\begin{figure}
  \includegraphics[height=8cm]{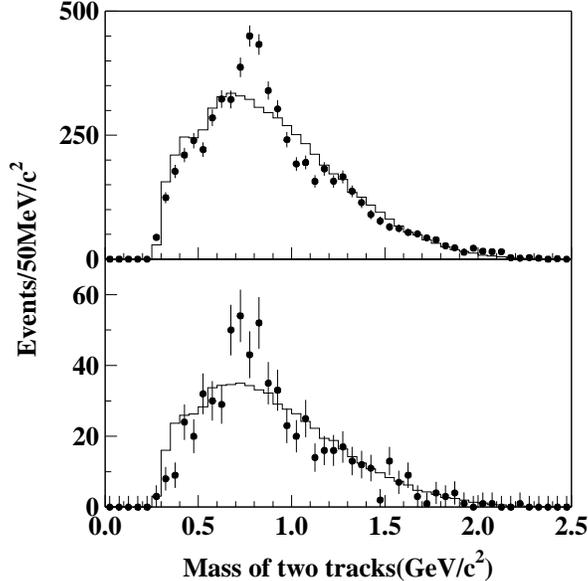}
\caption{\label{fig:6pires} The $\pp$ invariant mass distributions
  after applying all selection criteria.
  The upper plot is for $\psipto 3(\pp)$, while
  the lower one is for $\ee \rarr 3(\pp)$. The dots with error bars
  are data, while the histograms are MC simulated phase space
  distributions. All $\pp$ combinations are included.}
\end{figure}

\item Above, we assumed $\mathcal{F}(s) \propto 1/s$
  when subtracting the continuum contribution. Assuming
  a different dependency, such as $\mathcal{F}\propto 1/s^2$, yields a
  difference of 4\% and is regarded as the systematic
  error of the continuum subtraction.
  
\item The systematic uncertainty from the kinematic fit is estimated
  to be around 4\% by using a different MDC wire resolution simulation
  model~\cite{simbes}. This is in agreement with various studies using
  pure data samples which can be selected without using a kinematic
  fit~\cite{jpsi3pi}.

\end{enumerate}

Compared to the previous Mark-I result~\cite{mark1psp} of
$\mathcal{B}[\psipto 3(\pp)]=(1.5 \pm 1.0) \times 10^{-4}$, our
measurement has much better precision and a considerably higher
central value.

\subsection{\boldmath Form Factor of $\ee \rarr 3(\pp)$}
Using our values of $N^{obs}_{3.650}$ and $\mathcal{L}$ and
Eq.~\ref{born1}, the form factor at $\sqrt{s}=3.650$~GeV can be
determined.
Following the procedure in Ref.~\cite{ff}, assuming only one photon
annihilation in $\psipto 3(\pp)$ decays, one can also determine the
form factor at $\sqrt{s}=3.686$~GeV, using \cite{ff}
\begin{eqnarray}
  \sigma_{Born}(s)&=&\frac{4\pi\alpha^{2}}{3s}\times
  |\mathcal{F}(s)|^{2}\times[1+2\mathcal{R}B(s)+|B(s)|^2]\nonumber\\
  &\approx&\frac{4\pi\alpha^{2}}{3s}\times|\mathcal{F}(s)|^{2}\times[1+|B(s)|^2],
  \label{born2}
\end{eqnarray}
with
\begin{eqnarray*}
  B(s) &=& \frac{3\sqrt{s}\Gamma_{ee}/\alpha}{s-M_{\psip}^{2}+iM_{\psip}\Gamma_{t}},
\end{eqnarray*}
where $M_{\psip}$ is the mass of $\psip$ and $\Gamma _{ee}$ and
$\Gamma_{t}$ are the partial width to $\ee$ and total width of
$\psip$. In Eq.~(\ref{born2}), the interference term is neglected
since it is at the 1.3\% level~\cite{ff} and small compared with the
experimental uncertainties.

Using the numbers listed in Table~\ref{tab:val}, we obtain
\begin{eqnarray*}
  |\mathcal{F}(s=3.650^2)| &=&0.21 \pm 0.02,\\
  |\mathcal{F}(s=3.686^2)| &=&0.20 \pm 0.01.
\end{eqnarray*}

The form factors measured at the two energy points are consistent with
each other within $1\sigma$, i.e., there is no strong evidence for a
large contribution from other than the electromagnetic interaction
in $\psip$ decays.

Using the form factor measured at the $\psip$ peak, the branching
fraction of $\psipto 3(\pp)$ is determined to be:
\begin{equation}
  \mathcal{B}[\psipto 3(\pp)]=(4.8\pm 0.6) \times 10^{-4},
  \nonumber
\end{equation}
in good agreement with the result obtained by subtracting the
continuum contribution determined from the continuum data, but with
slightly improved precision since an extra assumption is introduced.
In the above calculations, $|\mathcal{F}(s)|\propto \frac{1}{s}$ is
assumed; assuming $|\mathcal{F}(s)|\propto \frac{1}{s^2}$ results in a
difference in $\mathcal{B}[\psipto 3(\pp)]$ less than 4\%, which is
taken as a systematic error in this branching fraction.

\section{\boldmath Measurement of $\jpsito 2(\pp)$}

As has been shown in Fig.~\ref{fig:mpp}, $\psipto 3(\pp)$ final
states can be used to measure the branching fraction of $\jpsito
2(\pp)$ via $\psipto \pp \jpsi, \jpsito 2(\pp)$.

In order to decrease the systematic error, we determine the
branching fraction of $\jpsito 2(\pp)$ from a comparison of the
following two processes as has been done in Ref.~\cite{jpsi3pi}:
\begin{itemize}
\item $\psipto\pp\jpsi,\jpsito 2(\pp)$\hfil (I)
\item $\psipto\pp\jpsi,\jpsito \mm$\hfil    (II)
\end{itemize}

The branching fraction is determined from
\begin{equation}
  \mathcal{B}[\jpsito 2(\pp)] =\frac{\displaystyle N^{obs}_{I}/\varepsilon_{I}}
  {\displaystyle N^{obs}_{II}/\varepsilon_{II}}\times \mathcal{B}(\jpsito\mm),
\label{b4pi}
\end{equation}
where $N^{obs}$ is the number of observed events, and
$\varepsilon$ is the efficiency. The branching fraction for the
leptonic decay $J/\psi\rightarrow \mu^+\mu^-$, is obtained from
the PDG~\cite{pdg}.

\subsection{Event Selection}
For process I, we use the same selection criteria as for $\psipto
3(\pp)$ except the $\jpsi$ and $\kzs \kzs$ vetos are no longer used.
To remove $\psipto \pp \jpsi,\jpsito K^* \overline{K} + c.c. \rarr
\kzs K^+\pi^- + c.c.$ background, events are vetoed if the recoil mass
of one $\pp$ pair falls into $(3.07,3.12)$ GeV$/c^{2}$ and, at the
same time, the mass of another $\pp$ pair falls into
$(0.47,0.53)$~GeV$/c^{2}$.

\begin{figure}
  \includegraphics[height=7cm]{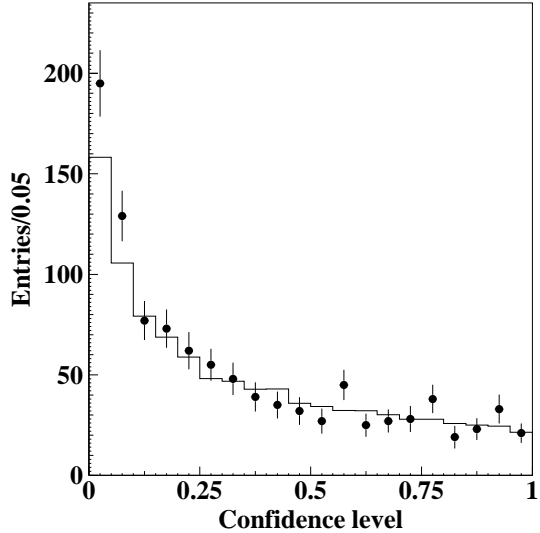}
\caption{\label{fig:x2_jpsi} The confidence level distribution for
  kinematic fitting of $\psipto \pp \jpsi, \jpsito 2(\pp)$ events. The
  dots with error bars are data, while the blank histogram is MC
  simulated signal.}
\end{figure}

Fig.~\ref{fig:x2_jpsi} shows confidence level distributions for the
kinematic fitting of $\psipto \pp \jpsi, \jpsito 2(\pp)$. The
agreement between data and MC simulation is very similar to that of
direct $\psipto 3(\pp)$ in Fig. \ref{fig:x2}(a). Because of the clean
$\jpsi$ signal seen below, we may conclude that there is nearly no
background in this process and that the discrepancy at small confidence
level is due to the simulation of the error matrix in track fitting,
which will be taken into consideration in the systematic error.

For process II, the selection criteria are similar to those in Ref.
\cite{jpsitoll}. The two lower momentum tracks are assumed to be
$\pp$, while the two higher momentum tracks $\mm$. The recoil mass of
$\pp$ candidates must fall within $(3.0,3.2)$ GeV$/c^{2}$, while the
invariant mass of $\mm$ candidates must be within 250 MeV$/c^{2}$ of
the $\jpsi$ mass. For $\mm$ candidates, each track must have $N^{hit}
\ge 2$, where $N^{hit}$ is the number of muon identification (MUID)
layers with matched hits and ranges from 0 to 3, indicating not a
muon(0), a weakly(1), moderately(2), or strongly(3) identified
muon~\cite{muid}.

\begin{figure}
  \includegraphics[height=7cm]{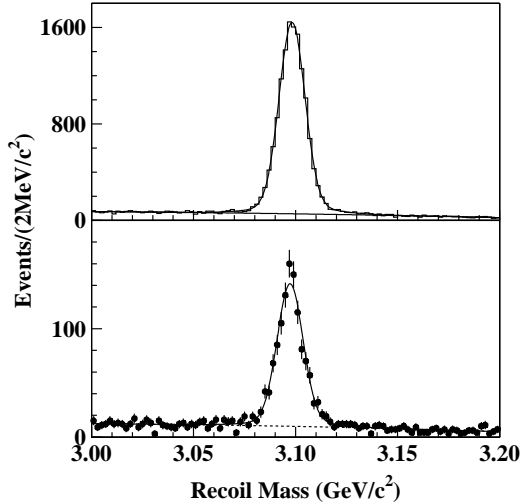}
\caption{\label{fig:Mppr} $\pp$ recoil mass distributions of
  $\psipto 3(\pp)$ candidates. The histogram is MC simulation (upper
  plot), while the dots with error bars are data (lower plot). The
  smooth curves show the best fits to the distributions as described
  in the text. All $\pp$ combinations are included, and the background
  mainly comes from the incorrect combinations.}
\end{figure}

The recoil mass distribution of all $\pp$ combinations of
$\psipto 3(\pp)$ candidate events is shown in Fig. \ref{fig:Mppr}.
Both data and MC simulation are fitted using a double-Gaussian for the
$\jpsi$ signal and a second order polynomial for the background.

\begin{figure}
  \includegraphics[height=7cm]{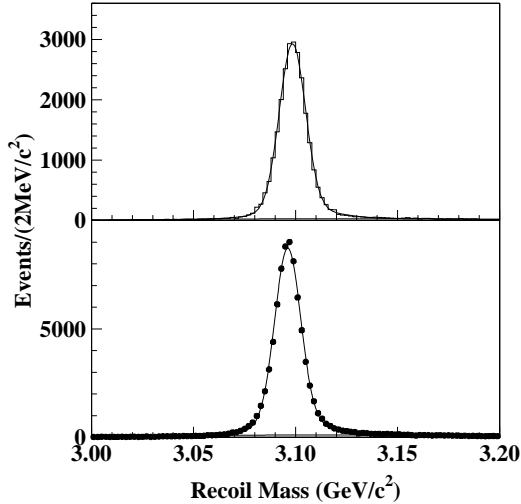}
\caption{\label{fig:Mppr.mu}The $\pp$ recoil mass distributions
  of $\psipto \pp \mm$ candidates. The histogram is MC simulation (upper plot),
  while the dots with error bars are data (lower plot). The smooth
  curves show the best fits to the distributions as described in the
  text.}
\end{figure}

Fig.~\ref{fig:Mppr.mu} shows the $\pp$ recoil mass distribution of
the $\psipto \pp\mm$ candidates. A similar fit is performed as for
the $3(\pp)$ mode.

From the fits we get $N^{obs}_{I}/N^{obs}_{II} = 0.027\pm 0.001$ and
$\varepsilon_{I}/\varepsilon_{II} = 0.46\pm 0.01$.

\subsection{Branching Fraction and Systematic Error}

Using Eq.~\ref{b4pi}, the relative branching fraction of $\jpsito
2(\pp)$ is determined to be:
\begin{equation}
  \frac{\displaystyle \mathcal{B}[\jpsito 2(\pp)]}
  {\displaystyle \mathcal{B}[\jpsito \mm]}
  =(5.86 \pm 0.19 \pm 0.39)\%, \nonumber
\end{equation}
where the first error is statistical and the second systematic.

Using the branching fraction of $\jpsito \mm$ from the PDG
\cite{pdg}, we obtain
\begin{equation}
  \mathcal{B}[\jpsito 2(\pp)]
  =(3.45 \pm 0.12 \pm 0.24) \times 10^{-3}. \nonumber
\end{equation}

Since this is a relative measurement, many systematic errors cancel,
either completely or partially. The remaining systematic errors are
listed in Table~\ref{tab:err2} and are described below.

\begin{table}
\caption{\label{tab:err2} Summary of systematic errors for
$\mathcal{B}[\jpsito 2(\pp)]$.}
  \begin{tabular}{l|c}\hline
    Source                &systematic errors (\%)\\\hline\hline
    MC Statistics         &1\\
    MDC tracking          &4\\
    Generator             &3\\
    Signal fitting        &1\\
    kinematic fitting     &4\\
    Muon identification   &2\\
    Background            &1\\
    Total error           &7\\\hline
  \end{tabular}
\end{table}

\begin{enumerate}
\item For process I, the statistical error of the MC sample is 0.9\%,
  while for process II, it is 0.7\%, including any
  uncertainties introduced by the fits.

\item There are two more tracks in $\jpsito 2(\pp)$ than in $\jpsito
  \mm$, the uncertainty in tracking is dominated by
  the two extra tracks and is estimated to be about 4\%.

\item Fig.~\ref{fig:jpsires} shows the $\pp$ mass distributions
after all the cuts, where clear $\rho$ and $f_2(1270)$ signals can be
seen. To estimate the efficiency difference between data and
MC simulation (pure phase space), we generate samples with
different intermediate states in $\jpsito 2(\pp)$, e.g., $\rho$,
$f_{2}(1270)$, $a_{2}(1320)$, etc. and find the differences
from the pure phase space MC simulation are about 3\%, which is
taken as the systematic error for the generator.

\begin{figure}
  \includegraphics[height=8cm]{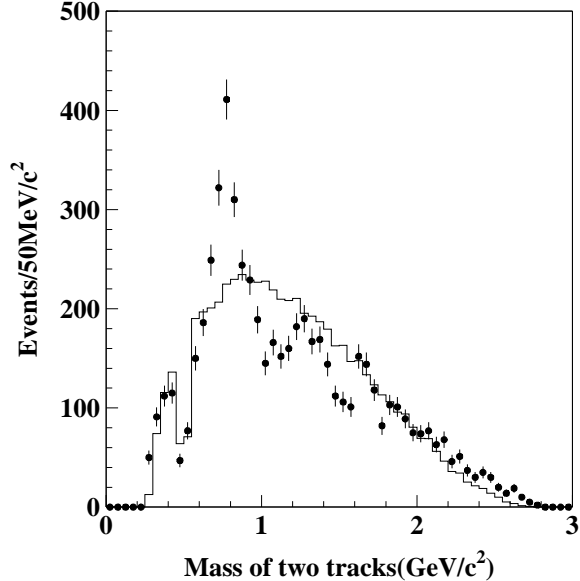}
\caption{\label{fig:jpsires} The $\pp$ invariant mass distributions
  for $\jpsito 2(\pp)$ candidates after applying all the selection
  criteria.  The dots with error bars are data, while the histogram is
  MC simulated phase space. The dip at the left side of
  the $\rho$ peak is due to the $\kzs$ mass cut. There are four entries
  for each $\jpsito 2(\pp)$ event.}
\end{figure}

\item The numbers of events obtained from the fits are affected by the
  background shape, as well as the signal shape. Using a linear
  background yields a change of the branching fraction of $1\%$, while
  using direct counting methods to estimate the numbers of events and
  efficiency results in negligible change in the branching fraction.
  The systematic error due to the fitting procedure is taken as $1\%$.

\item The systematic uncertainty from the kinematic fit is estimated
  the same as that of $\psipto 3(\pp)$ and is about $4\%$.
  
\item Instead of using the MUID, a $\mm$ sample may be obtained by
  using the energy deposited in the shower counter to remove $\jpsito
  e^+ e^-$ events and neglecting $\jpsito \pi^+ \pi^-$ and $K^+ K^-$
  which have small branching ratios~\cite{jpsi3pi}. From the number of
  events obtained with this method, we find that the error associated
  with the MUID requirement is about $2\%$.
  
\item Background contamination is
  determined using Monte Carlo simulation. The background
  fractions in both processes are far less than 1\%, and the uncertainty
  of the estimation is at 0.5\% level.
\end{enumerate}

Adding the systematic errors in quadrature, the total systematic error for
$\mathcal{B}[\jpsito 2(\pp)]$ is 7\%.

Our measurement of $\mathcal{B}[\jpsito 2(\pp)]$ is in good agreement
with the existing measurement by Mark-I~\cite{mark1jpsi}, which is
$(4.0\pm 1.0)\times 10^{-3}$, based on 76 observed events. Although
non resonance continuum $2(\pp)$ production was
mentioned in the Mark-I paper, it is not clear whether it was removed
in evaluating the $\jpsi$ decay branching fraction. Our result has
better precision compared with the Mark-I
result~\cite{mark1jpsi}. A recent measurement of this branching
fraction using initial state radiation events was reported by the BABAR
experiment~\cite{babar}; the result ($(3.61\pm 0.26\pm 0.26)\times
10^{-3}$) agrees with our measurement well but has a larger
error.

\section{Summary}

Using 14~M $\psip$ events and 6.42~pb$^{-1}$ of continuum data at
$\sqrt{s}=3.650$~GeV, the branching fractions of $\psipto 3(\pp)$ and
$\jpsito 2(\pp)$ are determined to be $(4.83\pm 0.38\pm 0.69) \times
10^{-4}$ and $(3.45\pm 0.12\pm 0.24)\times 10^{-3}$, respectively. The
former is larger than the Mark-I result using $\psip$
decays~\cite{mark1psp}, while the latter is consistent with Mark-I
measurement~\cite{mark1jpsi} using a $\jpsi$ decay sample.  In both cases,
our measurements have improved precision.

Using the above results and $\mathcal{B}[\psipto 2(\pp)]$,
$\mathcal{B}[\jpsito 3(\pp)]$ from PDG~\cite{pdg}, we obtain
\begin{eqnarray*}
  Q_{h} \equiv \frac{\displaystyle {\mathcal{B}}[\psipto 3(\pp)]}
  {\displaystyle {\mathcal{B}}[\jpsito 3(\pp)]} = (12 \pm 7)\%,\\
  Q_{h} \equiv \frac{\displaystyle {\mathcal{B}}[\psipto 2(\pp)]}
  {\displaystyle {\mathcal{B}}[\jpsito 2(\pp)]} = (13 \pm 3)\%.
\end{eqnarray*}
They are consistent with the ``$12\%$ rule"~\cite{apple}
expectation within errors.

\begin{acknowledgments}

The BES collaboration thanks the staff of BEPC for their hard
efforts and the members of IHEP computing center for their helpful
assistance. This work is supported in part by the National Natural
Science Foundation of China under contracts Nos. 19991480,
10225524, 10225525, the Chinese Academy of Sciences under contract
No. KJ 95T-03, the 100 Talents Program of CAS under Contract Nos.
U-11, U-24, U-25, and the Knowledge Innovation Project of CAS
under Contract Nos. U-602, U-34 (IHEP); by the National Natural
Science Foundation of China under Contract No. 10175060 (USTC),
and No. 10225522 (Tsinghua University); and by the U.S. Department of
Energy under Contract No. DE-FG02-04ER41291 (U of Hawaii).
\end{acknowledgments}

\end{document}